\documentclass[12pt]{article}

\usepackage{latexsym}

\usepackage{graphics}
\usepackage{graphicx}
\usepackage{epstopdf}

\begin{document}


\title{Charged Rotating Dilaton Black Holes with Kaluza-Klein Asymptotics}
\author{
   Christian Knoll$^{1}$, Petya Nedkova$^{1,2}$\thanks{E-mail:pnedkova@phys.uni-sofia.bg} \\
{\footnotesize ${}^{1}$Institut f\"{u}r Physik, Universit\"{a}t Oldenburg, D-26111 Oldenburg, Germany,}\\
{\footnotesize ${}^{2}$ Department of Theoretical Physics,
                Faculty of Physics, Sofia University,}\\
{\footnotesize  5 James Bourchier Boulevard, Sofia~1164, Bulgaria }\\
}

\date{}

\maketitle

\begin{abstract}
We construct a class of stationary and axisymmetric solutions to the 5D Einstein-Maxwell-dilaton gravity,  which describe configurations of
charged rotating black objects with Kaluza-Klein asymptotics. The solutions are constructed by uplifting a vacuum seed solution to six dimensions,
performing a boost, and a subsequent circle reduction. We investigate the physical properties of the charged solutions, and obtain their
general relations to the properties of the vacuum seed. We also derive the gyromagnetic ratio and the Smarr-like relations. As particular cases
we study three solutions, which describe a charged rotating black string, a charged rotating black ring on Kaluza-Klein bubbles, and a superposition of
two black holes and a Kaluza-Klein bubble.
\end{abstract}

\section{Introduction}

Einstein-Maxwell-dilaton gravity arises as a truncation of the low energy string theory action and has attracted a lot of attention as a quantum gravity
motivated gravitational theory.  A major topic of interest is the construction of black hole solutions within the theory in various spacetime dimensions
and the investigation of their properties. Static exact black hole solution are known in four dimensions \cite{Gibbons:1982}, \cite{Horowitz},
due to the integrability of the Einstein-Maxwell-dilaton gravity in this case. The dimensionally reduced action describes a symmetric target space
with a $SL(2,R)\times R$ hidden symmetry group which opens a way to the generation of new solutions by algebraic means \cite{Galtsov:1994}, \cite{Galtsov:1995}.
Thus, a dilaton-Weyl class of solutions can be constructed generalizing the static Einstein-Maxwell solutions
in the presence of a dilaton field. In the stationary case no nontrivial hidden symmetries are discovered
for general value of the dilaton coupling constant,  which  hinders the construction of exact solutions describing rotating charged dilaton black holes.
Two particular values of the dilaton constant $\alpha$ make an exception, when the hidden symmetry group is enhanced: $\alpha=0$ corresponding to the
Einstein-Maxwell case, and $\alpha = \sqrt{3}$, which is equivalent to the Kaluza-Klein theory.

The situation is similar in five dimensions, where the static
truncation of Einstein-Maxwell-dilaton theory is highly symmetric and allows for the construction of exact solutions, while in the stationary case the
integrability is violated. Even in the Einstein-Maxwell limit $\alpha=0$ no nontrivial hidden symmetries are discovered when the electromagnetic field is
fully exited \cite{Yazad:2011}. An integrable sector was found \cite{Yazad:2006a},\cite{Yazad:2006b}, in which the electromagnetic and the twist potentials
decouple and give rise to two independent sigma-models. Consequently, some stationary black hole solutions were constructed carrying magnetic (dipole)
charge \cite{Yazad:2007},\cite{Yazad:2008}.  However, electrically charged rotating solutions do not belong to this sector, as well as rotating solutions
with dipole charge, for which the electromagnetic vector potential has a component along some of the axes of rotation.

The only case when the dimensionally reduced action for a general stationary and axisymmetric electromagnetic field describes a symmetric space corresponds
to the Kaluza-Klein theory. It is contained as a particular case in the 5D Einstein-Maxwell-dilaton gravity for a dilaton coupling constant
 $\alpha =\sqrt{8/3}$. Thus, Kaluza-Klein solutions provide examples for exact solutions describing charged rotating black holes,
which do not have counterparts in the pure Einstein-Maxwell case. Due to the lack of solution generation techniques,
5D charged rotating black holes for dilaton coupling $\alpha\neq \sqrt{\frac{3}{8}}$ are investigated numerically
\cite{Kunz:2005}-\cite{Kunz:2013}, or perturbatively \cite{Kunz:2010a}, \cite{Kunz:2010b}. Some exact solutions exist only for
extremal charged rotating black holes with Kaluza-Klein asymptotics \cite{Matsuno:2012}.

In five dimensions a great variety of exact stationary and axisymmetric black hole solutions in vacuum were constructed. In addition to the higher
dimensional generalization of the Kerr solution \cite{Myers:1986},  asymptotically flat black objects with non-spherical horizon topology were discovered,
such as black rings \cite{Emparan:2002a}, and black lenses \cite{Teo:2008}. Non-linear superpositions of such objects were obtained resulting in different
multi-horizon solutions, including system of concentric or orthogonal black rings \cite{Elvang:2008}, \cite{Iguchi:2007},
and black Saturns \cite{Elvang:2007}. Black hole solutions with Kaluza-Klein asymptotics ($M^4\times S^1$) were also considered  motivated by brane world
scenarios with compact extra dimensions. They include a large class of solutions called black holes on Kaluza-Klein bubbles \cite{Emparan:2002}-
\cite{Nedkova:2010},
which represent superposition between black objects with different horizon topology and Schwarzschild instantons \cite{Hawking:1977}.
These systems can be further generalized to solutions which are only locally asymptotically Kaluza-Klein, but globally the spacelike boundary at infinity
represents some nontrivial fibration of $S^3$. In this way configurations of black objects and more complicated types of gravitational instantons were
obtained \cite{Chen:2011}. Different versions of such solutions including electromagnetic fields were constructed \cite{YN2}-\cite{Stelea:2013}.
They possess interesting thermodynamical properties, since certain electromagnetic fluxes also give contributions in their Smarr-like relations and
the first law of thermodynamics \cite{YN1}-\cite{Nedkova:2012}.

Charged generalizations of rotating vacuum solutions can be obtained in the Kaluza-Klein sector of Einstein-Maxwell-dilaton gravity by means of a simple
procedure. The $D$ dimensional vacuum solution is trivially embedded in $D+1$ dimensional spacetime by adding an extra dimension, a boost is performed with respect to the extra
dimension, and the resulting solution is dimensionally reduced again to $D$ dimensions along the additional coordinate. As a result a generalization of
the initial solution possessing an electromagnetic and a dilaton field is constructed. Some charged rotating dilaton black objects were obtained in five
dimensions in this way. They include the charged counterpart of the Myers-Perry black hole \cite{Kunz:2006a}, the black ring \cite{Kunduri:2004}, and the
black Saturn \cite{Grunau:2014}. More general black ring solutions with double rotation and carrying also a dipole charge were obtained by a related scheme
\cite{Rocha:2012}-\cite{Pomeransky:2012}. A suitable vacuum solution with a compact dimension is constructed in six dimensions by means of the inverse
scattering method. Performing a Kaluza-Klein reduction along the compact dimension leads to a solution to the 5D Enstein-Maxwell-dilaton gravity which can
possess a dipole charge \cite{Rocha:2012}, \cite{Chen:2012},  and an electric charge \cite{Pomeransky:2012}.

In this paper we obtain charged rotating solutions to the 5D Einstein-Maxwell-dilaton gravity, which describe configurations of black objects
with Kaluza-Klein asymptotics. We consider a class of stationary and axisymmetric solutions which contain at least one horizon
and for which the Killing vector associated with the compact dimension is hypersurface orthogonal.
Consequently, the solutions possess a single angular momentum, and the possible fixed point sets of the spacelike Killing fields
are restricted to Kaluza-Klein bubbles, and the axis of rotation. Such solutions describe configurations of black objects with different topology,
which may be also superposed with Kaluza-Klein bubbles. Given a particular vacuum configuration, we apply on it the described Kaluza-Klein transformation
by uplifting it to six dimensions, performing a boost, and a subsequent circle reduction. As a result we obtain the corresponding charged dilaton
solution in the Kaluza-Klein sector of the 5D Einstein-Maxwell-dilaton gravity. Various general relations exist between the properties of the vacuum seed
solution and the charged dilaton one. We investigate these relations and show how the physical characteristics of the charged solution can be expressed
by means of the characteristics of its vacuum counterpart.

The paper is organized in the following way. In sections 2 and 3 we describe the class of solutions we consider, and the solution generation technique.
In section 4 we study the physical properties of the charged solutions and their relations to the properties of the vacuum ones. In particular we
consider the conserved charges, the local Komar masses and angular momenta of the black holes, the local electric charges associated with the horizons,
the Smarr-like relations, and the gyromagnetic ratio. In section 5 we apply our results for the investigation of three particular solutions -  the charged
rotating black string, the charged rotating black ring on Kaluza-Klein bubbles, and a superposition of two rotating black holes and a Kaluza-Klein bubble.
In the paper we work in geometrical units $G=c=1$.

\section{Class of rotating black holes with Kaluza-Klein asymptotics}

In this work we consider a class of stationary and axisymmetric solutions with Kaluza-Klein asymptotics, which contain at least one horizon. The solutions
possess $R\times U(1)^2$ isometry group generated by three Killing vectors, one of which is asymptotically timelike. We can introduce a coordinate system
adapted to their orbits, such that the asymptotically timelike Killing vector is expressed as $\xi = \frac{\partial}{\partial t}$, the Killing vector
associated with the compact dimension is given by $k = \frac{\partial}{\partial \psi}$, and the remaining spacelike Killing vector is
$\eta = \frac{\partial}{\partial \phi}$. We consider only solutions for which the Killing vector $k$ is hypersurface orthogonal. Such solutions rotate
only with respect to the axis of the Killing vector $\eta$, and do not contain fixed point sets of nontrivial linear combinations of the spacelike Killing
vectors. The general form of the metric for this class is given by

\begin{eqnarray}
ds^2=g_{tt}(dt + \omega d\phi)^2 + {\tilde g}_{\phi\phi}d\phi^2 + g_{\psi\psi}d\psi^2 + \gamma_{ab}dx^a dx^b, \quad~~~ a,b = 1,2,
\end{eqnarray}
where $x^a, a=1,2$ are the coordinates parameterizing the 2-dimensional surfaces orthogonal to the Killing fields, and all the metric functions depend only
on them. A useful coordinate system are the canonical Weyl coordinates $\rho$ and $z$, in which the 2D metric is diagonal, and the determinant of the 3D
metric spanned by the Killing fields is equal to $-\rho^2$.

A convenient way to classify the 5D stationary and axisymmetric solutions is by means of their interval structure \cite{Hollands:2007}. The interval
structure gives information about the existence and the position of the horizons and the different fixed point sets of spacelike Killing fields. We consider the factor space of the spacetime with respect to the isometry group $\hat{M} = M/ R\times U(1)^2$. In Weyl coordinates it is diffeomorphic to the upper half-plane $\rho>0$, and all the horizons and the fixed point sets of the spacelike Killing vectors are located on its boundary $\rho=0$. Consequently, they correspond to finite intervals on the $z$-axis, given by  $z_i \leq z \leq z_{i+1}$ for some values $z_i$ and  $z_{i+1}$, or semi-infinite intervals on the $z$-axis. We can associate with each interval a direction vector, which specifies the linear combination of Killing vectors which vanishes on it, or equivalently, the particular kind of fixed point set which it describes. The set of data, containing the position and the type of the intervals for a given solution, and their lengths and directions is called interval structure.

\begin{figure}[htp]
\setlength{\tabcolsep}{ 0 pt }{\scriptsize\tt
		\begin{tabular}{ cc }
	\hspace{1.5cm} \includegraphics[width=12cm]{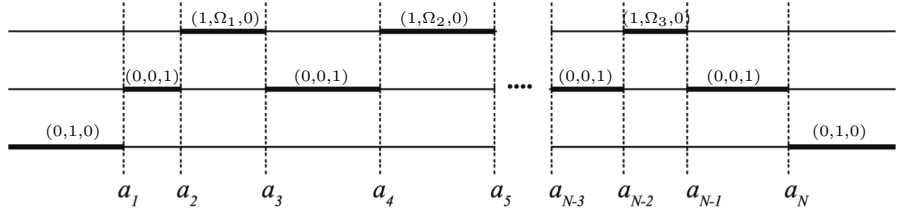}
                  \end{tabular}}
           \begin{picture}(0,0)(0,0)
\put(-300,12){{\tiny(0,0,1)}}
\put(-330,-9){{\tiny(0,1,0)}}
\put(-275,34){{\tiny(1,$\Omega_1$,0)}}
\put(-236,12){{\tiny(0,0,1)}}
\put(-196, 34){{\tiny(1,$\Omega_2$,0)}}
\put(-136,12){{\tiny(0,0,1)}}
\put(-112, 34){{\tiny(1,$\Omega_3$,0)}}
\put(-80,12){{\tiny(0,0,1)}}
\put(-40,-9){{\tiny(0,1,0)}}
\end{picture}
           \caption{\footnotesize{Interval structure of sequence of rotating black holes and Kaluza-Klein bubbles.}}
           \label{rodstr}
\end{figure}

As an example, the interval structure of a solution belonging to the class we consider is presented in fig. \ref{rodstr}.  The parameters $a_i$ correspond to the
endpoints of the intervals on the $z$-axis. The direction vector is specified over each interval, and it gives the coefficients in the linear combination
vanishing on it, with respect to a basis of Killing vectors $\{\frac{\partial}{\partial t}, \frac{\partial}{\partial \phi}, \frac{\partial}{\partial \psi}\}$.
The finite intervals correspond either to horizons (with direction $(1, \Omega_i, 0)$), or to fixed point sets of the Killing field associated with the
compact dimension (with direction $(0,0,1)$). These fixed point sets form regular 2D surfaces, which are called Kaluza-Klein bubbles. Consequently,
the presented interval structure describes a sequence of alternating black holes and Kaluza-Klein bubbles.  Each horizon rotates with angular velocity
$\Omega_i$ around the axis of the Killing field $\frac{\partial}{\partial\phi}$. The semi-infinite intervals with direction $(0,1,0)$ represent the axis
of rotation.

The most general interval structure associated to a solution within the class we consider contains three types of intervals: finite intervals with
direction $(1, \Omega_i, 0)$, corresponding to rotating black objects with angular velocity $\Omega_i$, finite intervals with direction $(0,0,1)$
corresponding to Kaluza-Klein bubbles, and finite and semi-infinite intervals with direction $(0,1,0)$ representing the axis of rotation. Both
semi-infinite intervals should be directed along the Killing vector $\eta= \frac{\partial}{\partial\phi}$ in order for the solutions to have
Kaluza-Klein asymptotics.

\section{Construction of the charged dilaton solutions}

We consider the Einstein-Maxwell-dilaton gravity in five dimensions with the action

\begin{equation}
I = {1\over 16\pi} \int d^5x \sqrt{-g}\left(R - 2g^{\mu\nu}\partial_{\mu}\varphi \partial_{\nu}\varphi  -
e^{-2\alpha\varphi}F^{\mu\nu}F_{\mu\nu} \right).
\end{equation}
It leads to the  following field equations
\begin{eqnarray} \label{FE}
R_{\mu\nu} &=& 2\partial_{\mu}\varphi \partial_{\nu}\varphi + 2e^{-2\alpha\varphi} \left[F_{\mu\rho}F_{\nu}^{\rho} - {1\over 6}g_{\mu\nu} F_{\beta\rho}
F^{\beta\rho}\right], \\
\nabla_{\mu}\nabla^{\mu}\varphi &=& -{\alpha\over 2} e^{-2\alpha\varphi} F_{\nu\rho}F^{\nu\rho}, \nonumber \\
&\nabla_{\mu}&\left[e^{-2\alpha\varphi} F^{\mu\nu} \right]  = 0, \nonumber
\end{eqnarray}
where $R_{\mu\nu}$ is the Ricci tensor for the spacetime metric $g_{\mu\nu}$, $F_{\mu\nu}$ is the Maxwell tensor, and
$\varphi$ is the dilaton field. We assume that the dilaton coupling parameter takes the value $\alpha = \sqrt{8/3}$,  which constrains the Kaluza-Klein
sector in the theory.

Solutions to the 5D Einstein-Maxwell-dilaton gravity with $\alpha = \sqrt{8/3}$ can be obtained by the following procedure. We consider a known vacuum
solution to the 5D Einstein equations,  which we call a seed. For our purposes, we assume that the seed solution has the general form
 \begin{eqnarray}\label{seed}
ds^2_{0}=g^{(0)}_{tt}(dt + \omega^{(0)} d\phi)^2 + {\tilde g}^{(0)}_{\phi\phi}d\phi^2 + g^{(0)}_{\psi\psi}d\psi^2 +
\gamma^{(0)}_{ab}dx^a dx^b,
\end{eqnarray}
i.e. it belongs to the class of solutions described in section 2.  We embed it trivially in six dimensions by adding an extra dimension. Thus, we obtain
a vacuum solution to the 6D Einstein equations with the metric

\begin{equation}
ds_{6}^2 =  ds_{0}^2 + dx^2_6 \ ,
\end{equation}
where $ds_{0}^2$ is the metric of the seed solution, and the coordinate $x_6$ parameterizes the extra dimension. Then, we perform a hyperbolic rotation in
the $t-x_6$ plane with the matrix

\begin{equation}
L=\left(
\begin{array}{cc}
\cosh\gamma&\sinh\gamma \\
\sinh\gamma&\cosh\gamma
\end{array}
\right) \ ,
\end{equation}
and reduce the resulting solution along the extra dimension using the ansatz

\begin{eqnarray}
ds^2_{6}=e^{\sqrt{\frac{2}{3}}\varphi} ds^2_{5} +
e^{-2\sqrt{\frac{3}{2}}\varphi}\left(dx_{6} +
2A_{\mu}dx^\mu\right)^2 .
\end{eqnarray}
The five-dimensional metric $ds^2_{5}$ corresponds to a solution in the Kaluza-Klein sector of the 5D Einstein-Maxwell-dilaton gravity with a scalar field
$\varphi$ and an electromagnetic potential $A_{\mu}$. The metric of constructed solution is connected to the seed solution $(\ref{seed})$ as

\begin{eqnarray}\label{metric}
ds^2_{5}= \Lambda^{-\frac{2}{3}}g^{(0)}_{tt}(dt + \cosh\gamma\,\omega^{(0)} d\phi)^2 + \Lambda^{\frac{1}{3}}\left[{\tilde g}^{(0)}_{\phi\phi}d\phi^2 +
 g^{(0)}_{\psi\psi}d\psi^2 + \gamma^{(0)}_{ab}dx^a dx^b \right],
\end{eqnarray}
where the function $\Lambda$ is given by
\begin{eqnarray}
\Lambda = \cosh^2\gamma + \sinh^2\gamma g^{(0)}_{tt},
\end{eqnarray}
while the Maxwell 2-form $F$ and the dilaton field are
\begin{eqnarray}\label{EM_field}
F &=& dt\wedge dA_{t} + d\phi\wedge dA_{\phi},  \\
A_{t} &=& \frac{1}{2}\Lambda^{-1}\sinh\gamma\cosh\gamma\, (1 + g^{(0)}_{tt}), \nonumber \\
A_{\phi}&=& \frac{1}{2}\Lambda^{-1}\sinh\gamma\,\omega^{(0)}g^{(0)}_{tt}, \nonumber \\
e^{-2\alpha\varphi} &=& e^{-4\sqrt{\frac{2}{3}}\varphi} = \Lambda^{\frac{4}{3}}. \nonumber
\end{eqnarray}

The constructed solution possesses the same interval structure as the seed solution, i.e. it contains the same number of disconnected horizons and
the same fixed point sets of the spacelike Killing vectors. The only difference is that the angular velocity of each horizon is modified by a factor
$\cosh^{-1}\gamma$. Thus, if the $i$-th horizon in the seed solution is rotating with  angular velocity $\Omega^{(0)}_i$, the angular velocity of the
same horizon in the charged solution will be $\Omega_i = \cosh^{-1}\gamma\Omega^{(0)}_i$.

Solutions with multiple disconnected fixed point sets of a spacelike Killing field may suffer from conical singularities. To avoid conical singularities
the orbits of the Killing field in the vicinity of each of its fixed point sets should have the same periodicity, which should be also consistent with
the asymptotic structure of the solution. For our class of solutions it follows that on each of the bubble intervals we should have $\Delta\psi = L$,
where $L$ is the length of the Kaluza-Klein circle at infinity, while on the axis of rotation is should be satisfied $\Delta\phi = 2\pi$. In Weyl
canonical coordinates these conditions possess the form

\begin{eqnarray}\label{conical}
(\Delta\psi)_{I_{B_j}} = 2\pi \lim_{\rho\to 0}\sqrt{{\rho^2 g_{\rho\rho}\over g_{\psi\psi}}}= L, \quad~~~
(\Delta\phi)_{I_{\eta}} = 2\pi \lim_{\rho\to 0}\sqrt{{\rho^2 g_{\rho\rho}\over g_{\phi\phi}}}= 2\pi,
\end{eqnarray}
where $I_{B_j}$ is the $j$-th bubble interval, and $I_{\eta}$ is an interval corresponding to the rotation axis. Considering the general form of the
metric $(\ref{metric})$  we see that the charging transformation does not influence the regularity conditions.\footnote{We do not consider solutions
which contain closed timelike curves.} Therefore, if the seed solution is free of conical singularities and the length of its Kaluza-Klein circle at
infinity is equal to $L^{(0)}$, the charged solution will be also free of conical singularities with the same length of the compact dimension at
infinity $L = L^{(0)}$. In the following discussion we will consider only balanced solutions.

\section{Physical Properties}

\subsection{Conserved charges}

The constructed class of solutions are characterized by a number of conserved charges - the ADM mass $M_{ADM}$ and angular momentum ${\cal J}$,
the electric charge ${\cal Q}$, and the dilaton charge ${\cal D}$. They can be computed by the following integrals

\begin{eqnarray}\label{MJ}
M_{ADM} = - {L\over 16\pi} \int_{S^{2}_{\infty}} \left[2i_k \star d\xi - i_\xi \star d k \right], \quad~~~
{\cal J} = \frac{L}{16\pi}\int_{S^{2}_{\infty}}  i_k \star d\eta,
\end{eqnarray}
\begin{eqnarray}\label{Q}
{\cal Q} = \frac{1}{4\pi}\int_{S^{2}_{\infty}\times S^1} e^{-2\alpha\varphi}\star F, \quad ~~~
{\cal D} = -\frac{1}{4\pi}\int_{S^{2}_{\infty}} i_ki_\xi\star d\varphi,
\end{eqnarray}
over the 2-dimensional sphere at spacial infinity $S^2_\infty$. Due to the Kaluza-Klein asymptotics, the solutions are characterized by an additional
conserved charge - the gravitational tension \cite{Traschen:2001}, \cite{Traschen:2003}. It can be calculated by the generalized Komar integral
\cite{Townsend:2001}, \cite{YN1}

\begin{eqnarray}\label{T}
{\cal{T}} =  - {1\over 16\pi} \int_{S^{2}_{\infty}} \left[i_k \star d\xi - 2i_\xi \star d k \right].
\end{eqnarray}
The conserved charges can be extracted from the asymptotic behavior of the metric functions, and the electromagnetic and dilaton potentials. We
introduce the asymptotic coordinates $r$ and $\theta$ defined as

\begin{eqnarray}
\rho=r\sin{\theta}, \quad~~~
z=r\cos{\theta}, \nonumber
\end{eqnarray}
and consider the general expansion of the metric functions at the asymptotic infinity $r \rightarrow \infty$
\begin{eqnarray}
g_{tt} &\approx& -1 + \frac{c_t}{r}, \quad~~~ g_{\psi\psi} \approx 1 + \frac{c_\psi}{r}.
\end{eqnarray}
Then, the Komar integrals for the ADM mass and the gravitational tension are equal to
\begin{eqnarray}
M_{ADM} = \frac{L}{4}( 2c_t - c_\psi ), \quad~~~
{\cal T}L = \frac{L}{4}(c_t - 2c_\psi ) .
\end{eqnarray}
The angular momentum, the electric charge and the dilaton charge are contained as coefficients in the leading order terms in the asymptotic expansion
of the metric function $g_{t\phi}$, the electromagnetic potential $A_t$, and the dilaton potential $\varphi$, respectively,

\begin{eqnarray}
g_{t\phi}&\approx& -\frac{2{\cal J}}{L} \frac{\sin^2\theta}{r}, \quad~~~ A_t \approx  \frac{{\cal Q}}{Lr}, \quad~~~ \varphi \approx -\frac{{\cal D}}{r}
\end{eqnarray}

We can also define the magnetic moment ${\cal M}$ as the coefficient in the asymptotic expansion of the electromagnetic potential $A_\phi$
\begin{eqnarray}
A_{\phi} \approx - \frac{{\cal M}}{L}\frac{\sin^2\theta}{r}.
\end{eqnarray}
The conserved charges for our class of charged dilaton solutions can be expressed by the characteristics of the corresponding seed solution in the
following way. We obtain the asymptotic expansions  for the metric functions and the potentials in the form

\begin{eqnarray}
g_{tt} &\approx& -1 + \frac{c_t}{r}  = -1 + \frac{c^{(0)}_t}{r} + \frac{2}{3}\sinh^2\gamma\frac{c_t^{(0)}}{r},   \\[2mm]
g_{t\phi}&\approx&  -\cosh\gamma\frac{2{\cal J}^{(0)}}{L} \frac{\sin^2\theta}{r},  \nonumber\\[2mm]
g_{\psi\psi} &\approx& 1 + \frac{c_\psi}{r} = 1 + \frac{c^{(0)}_\psi}{r} + \frac{1}{3}\sinh^2\gamma\frac{c_t^{(0)}}{r} ,  \nonumber \\[2mm] \nonumber
\end{eqnarray}
\begin{eqnarray}
A_t &\approx& \frac{1}{2}\sinh\gamma\cosh\gamma\frac{c_t^{(0)}}{r},   \\[2mm]
A_{\phi}&\approx& -\sinh\gamma \frac{{\cal J}^{(0)}}{L} \frac{\sin^2\theta}{r},  \nonumber \\[2mm]
\varphi &\approx& - \frac{1}{\sqrt{6}}\sinh^2\gamma\frac{c_t^{(0)}}{r}, \nonumber
\end{eqnarray}
where we denote with zero index the corresponding quantities for the seed solution, and $L$ is the length of the Kaluza-Klein circle at infinity.
Consequently, the ADM mass, the angular momentum and the tension are related to the conserved charges of the seed solution as

\begin{eqnarray}
M_{ADM} &=& M^{(0)}_{ADM} + \frac{L}{4}\sinh^2\gamma c_t^{(0)}, \quad~~~ {\cal J} = \cosh\gamma{\cal J}^{(0)}, \quad~~~{\cal T} =  {\cal T}^{(0)},
\end{eqnarray}
while the electric charge,  the  dilaton charge and the magnetic moment are

\begin{eqnarray}
{\cal Q} = \frac{L}{2}\sinh\gamma\cosh\gamma c_t^{(0)}, \quad~~~ {\cal D} = \frac{1}{\sqrt{6}}\sinh^2\gamma c_t^{(0)},
\quad~~~ {\cal M} = \sinh\gamma {\cal J}^{(0)}.
\end{eqnarray}
The magnetic moment and the dilaton charge are not independent characteristics of the solutions.  They are related to the product of the angular momentum,
or electric charge, respectively, and the so called corotating potential $\Sigma_H$ defined on the black hole horizons. The corotating potential on the
$i$-th black hole horizon is defined as

\begin{eqnarray}
\Sigma_{H_i} = (v^\mu A_\mu)_{H_i},
\end{eqnarray}
where $A_\mu$ is the electromagnetic potential, and $v = \xi  + \Omega_i \eta$ is the Killing vector which becomes null on the corresponding horizon. For
the class of solutions, which we consider the corotating potential takes the constant value $\Sigma_H = \frac{1}{2}\tanh\gamma$ on every black hole horizon.
Thus, the dilaton charge and the magnetic moment are equal to
\begin{eqnarray}
{\cal D} = \sqrt{\frac{8}{3}}\frac{\Sigma_H{\cal Q}}{L} = \alpha \frac{\Sigma_H{\cal Q}}{L}, \quad~~~{\cal M} = 2\Sigma_H {\cal J}.
\end{eqnarray}

\subsection{Twist potentials and local quantities}

We consider the twist 1-form $i_\xi i_k \star d \xi$. It satisfies the Ricci identity

\begin{eqnarray}\label{chi}
d i_\xi i_k \star d \xi &=& 2\star\left[R(\xi)\wedge k \wedge \xi \right],
\end{eqnarray}
where $R(\xi)$ is the Ricci 1-form, and from the field equations we can obtain the identity

\begin{eqnarray}
\star R(\xi) = - 2e^{-2\alpha\varphi} \left( -{2\over 3}i_{\xi}F\wedge \star F + {1\over 3} F\wedge i_{\xi}\star F \right).
\end{eqnarray}
Using the explicit form of the electromagnetic field $(\ref{EM_field})$, it is reduced to

\begin{eqnarray}
\star\left[R(\xi)\wedge k \wedge \xi \right] = 2dA_t\wedge i_\xi
i_k e^{-2\alpha\varphi}\star F .
\end{eqnarray}

The field equations imply that $di_\xi i_k e^{-2\alpha\varphi}\star F = 0$, consequently we can introduce an electromagnetic potential ${\cal B}_\xi$ such
that $d{\cal B}_\xi = i_\xi i_k e^{-2\alpha\varphi}\star F$. Taking advantage of it, eqn. ($\ref{chi}$) yields
\begin{eqnarray}
d i_\xi i_k \star d \xi -  4d\left(A_t d{\cal B}_\xi\right)=0 .
\end{eqnarray}

The 1-form $i_\xi i_k \star d \xi - 4 A_t d{\cal B}_\xi$ is invariant under the Killing fields $ \xi,k$ and $\eta$ and can be viewed as defined on
the factor space $\hat{M}= M/ R \times U(1)^2$.  The factor space is simply connected, therefore
there exists a globally defined potential $\chi$, such that

\begin{equation}
 d\chi = i_\xi i_k \star d \xi - 4A_t d{\cal B}_\xi.
\end{equation}

In a similar way we can define a potential related to the twist 1-form $i_\eta i_k \star d \eta$  with respect to the Killing field $\eta$. It satisfies
the relation
\begin{eqnarray}
d i_\eta i_k \star d \eta &=& 2\star\left[R(\eta)\wedge k \wedge \eta \right],
\end{eqnarray}
where

\begin{eqnarray}
\star\left[R(\eta)\wedge k \wedge \eta \right] = 2dA_\phi\wedge i_\eta
i_k e^{-2\alpha\varphi}\star F .
\end{eqnarray}

Introducing an electromagnetic potential ${\cal{B_\eta}}$ such that $d{\cal{B_\eta}} = i_\eta
i_k e^{-2\alpha\varphi}\star F$, we obtain the identity

\begin{eqnarray}
d i_\eta i_k \star d \eta -  4d\left(A_\phi d{\cal B}_\eta\right)=0,
\end{eqnarray}
which implies that there exists a potential $\mu$ such that

\begin{equation}
 d\mu = i_\eta i_k \star d\eta- 4A_\phi d{\cal B}_\eta.
\end{equation}

We consider the corresponding potentials $\chi^{(0)}$ and $\mu^{(0)}$ for the vacuum seed solution, defined by the relations
$d\chi^{(0)} = (i_\xi i_k \star d \xi)^{(0)}$ and $d\mu^{(0)} = (i_\eta i_k \star d\eta)^{(0)}$ in terms of the corresponding Killing fields.
We introduce in addition the potential $\lambda^{(0)}$ satisfying $d\lambda^{(0)} = (i_\eta i_k \star d \xi)^{(0)}$. By direct calculation
we can prove that the following identities are satisfied

\begin{eqnarray}\label{twist_potential}
i_\xi i_k \star d \xi &=& \cosh\gamma\, d\chi^{0} - 2A_t\sinh\gamma\,d\chi^{0}, \\[2mm]
i_\eta i_k \star d\eta &=& \cosh\gamma d\mu^{(0)} - 2A_\phi\sinh\gamma\cosh\gamma d\lambda^{(0)}, \nonumber \\[2mm]
d{\cal B}_\xi &=& -\frac{1}{2}\sinh\gamma d\chi^{(0)}, \nonumber \\[2mm]
d{\cal B}_\eta &=& -\frac{1}{2}\sinh\gamma\cosh\gamma d\lambda^{(0)}, \nonumber
\end{eqnarray}

Consequently, we obtain the relations

\begin{eqnarray}
i_\xi i_k \star d \xi &=& \cosh\gamma\, d\chi^{0} + 4A_t d{\cal B}_\xi, \\[2mm]
i_\eta i_k \star d\eta &=& \cosh\gamma d\mu^{(0)} + 4A_\phi d{\cal B}_\eta,
\end{eqnarray}
and the twist potentials are connected as

\begin{eqnarray}
d\chi = \cosh\gamma d\chi^{(0)},  \quad~~~ d\mu = \cosh\gamma d\mu^{(0)}.
\end{eqnarray}

The identities ($\ref{twist_potential}$) lead to relations between some local characteristics of the charged dilaton solution and the corresponding
vacuum seed solution. We can define an intrinsic mass $M_{H_i}$ and angular momentum $J_{H_i}$ of each black hole by evaluating the integrals $(\ref{MJ})$
on  its horizon. Using that the Killing field $v = \xi + \Omega_i\eta = \xi - \omega^{-1}_i\eta $ vanishes on the horizon $H_i$ and that the corotating potential is constant on each
horizon, i.e. $\Sigma_H = \frac{1}{2}\tanh\gamma$, we obtain the following relations

\begin{eqnarray}\label{M_Komar}
M_{H_i} &=& -\frac{L}{16\pi}\int_{H_i} \left[2i_k \star d\xi - i_\xi \star d k\right] = \frac{L}{4}\int_{I_{H_i}} i_\eta i_k \star d\xi =\frac{L}{4}\omega_i\int_{I_{H_i}}
i_\xi i_k \star d\xi \nonumber \\[2mm]
&=&\frac{L}{4}\omega^{(0)}_i\int_{I_{H_i}}\cosh^2\gamma(i_\xi i_k \star d\xi)^{(0)} + L\,\omega_i\int_{I_{H_i}}A_t d {\cal B}_\xi \nonumber \\[2mm]
&=& \frac{L}{4}\omega^{(0)}_i\int_{I_{H_i}} (i_\xi i_k \star d\xi)^{(0)} - L\,\omega_i\int_{I_{H_i}}
\Sigma_{H}d {\cal B}_\xi + L\,\omega_i\int_{I_{H_i}}A_t d {\cal B}_\xi =   \nonumber \\[2mm]
 &=& M^{(0)}_{H_i} + L\int_{I_{H_i}}A_\phi d {\cal B}_\xi,
\end{eqnarray}
\begin{eqnarray}\label{J_Komar}
J_{H_i} &=& \frac{L}{16\pi}\int_{H_i} i_k \star d\eta =  -\frac{L}{8}\int_{I_{H_i}} i_\eta i_k \star d\eta =
 -\frac{L}{8}\int_{I_{H_i}}\cosh\gamma (i_\eta i_k \star d\eta)^{(0)} \nonumber \\[2mm]
 &-& \frac{L}{2}\int_{I_{H_i}}A_\phi d {\cal B}_\eta
  =\cosh\gamma J^{(0)}_{H_i} - \frac{L}{2}\int_{I_{H_i}}A_\phi d {\cal B}_\eta.
\end{eqnarray}
The integration is performed over the interval $I_{H_i}$ associated with the $i$-th horizon in the interval structure, and we denote by  $M^{(0)}_{H_i}$
and $J^{(0)}_{H_i}$  the intrinsic Komar mass and angular momentum of the corresponding black hole in the seed solution. We see that these quantities
are modified due to the interaction with the electromagnetic field. Similar behavior is observed, for example, for the Kerr-Newman black hole, or
magnetized black holes when the magnetic field is aligned with the axis of rotation (e.g. \cite{Galtsov:1986}).

 We can also define an intrinsic mass $M_{B_j}$ of each bubble by considering the Komar integral $(\ref{MJ})$ on its surface

\begin{eqnarray}
M_{B_j} &=& \frac{L}{16\pi}\int_{B_j}  i_\xi \star dk = -\frac{L}{8}\int_{I_{B_j}} i_\eta i_\xi \star dk .
\end{eqnarray}
The intrinsic mass is related to the length of the  interval $\Delta l_{B_j}$ associated with the bubble in the interval structure
\begin{eqnarray}
M_{B_j} &=& \frac{L}{4}\Delta l_{B_j} = M^{(0)}_{B_j}.
\end{eqnarray}
Consequently, it coincides with the intrinsic mass $M^{(0)}_{B_j}$ of the corresponding bubble in the seed solution, if the balance conditions
$(\ref{conical})$ are satisfied.

We can further assign a local charge ${\cal Q}_{H_i}$ of each black hole by evaluating the integral ($\ref{Q}$) on its horizon. Performing the
calculation

\begin{eqnarray}
{\cal Q}_{H_i} &=& \frac{1}{4\pi}\int_{H_i} e^{-2\alpha\varphi}\star F = -\frac{L}{2}\int_{I_{H_i}}i_\eta i_k e^{-2\alpha\varphi}\star F
=  - \frac{L}{2}\int_{H_i} d{\cal B}_\eta  \\[2mm] \nonumber
&=&\frac{L}{4}\sinh\gamma\cosh\gamma\int_{H_i}(i_\eta i_k \star d\xi )^{(0)} = \sinh\gamma\cosh\gamma M^{(0)}_{H_i},
\end{eqnarray}
we obtain that the local charge on each horizon is proportional to the intrinsic mass of the corresponding black hole in the seed solution. Then, the
total electric charge of the solution can be expressed as

\begin{eqnarray}
{\cal Q} = \sum_i {\cal Q}_{H_i} = \sinh\gamma\cosh\gamma\sum_i M^{(0)}_{H_i}.
\end{eqnarray}

Other quantities, which are of physical interest, are the  surface gravity $\kappa_{H_i}$ and the area $A_{H_i}$ of each horizon
\begin{eqnarray}
\kappa_{H_i} = \sqrt{-{1\over2}v_{\mu;\nu}v^{\mu;\nu}}|_{H_i}, \quad~~~
A_{H_i} = \int_{H_i} \sqrt{g_{H_i}}dz d\phi d\psi, \nonumber
\end{eqnarray}
where $v = \xi  + \Omega_i \eta$. Explicit calculation shows that they are related to the corresponding quantities for the seed solution as

\begin{eqnarray}
\kappa_{H_i} = \cosh\gamma^{-1} \kappa^{(0)}_{H_i}, \quad~~~ A_{H_i} = \cosh\gamma A^{(0)}_{H_i},
\end{eqnarray}
so that the charging transformation preserves their product. Taking into account the Smarr relation on each horizon

\begin{equation}\label{Smarr_H}
M_{H_i} = \frac{1}{4\pi}\kappa_{H_i}A_{H_i} + 2\Omega_iJ_{H_i},
\end{equation}
the horizon Komar masses and angular momenta should satisfy
\begin{equation}
M_{H_i} - 2\Omega_iJ_{H_i} = M^{(0)}_{H_i} - 2\Omega^{(0)}_iJ^{(0)}_{H_i},
\end{equation}
which is consistent with the derived expressions $(\ref{M_Komar})$-$(\ref{J_Komar})$.

\subsection{Smarr-like relations}

Let us consider the generalized Komar integral for the ADM mass $(\ref{MJ})$. It is convenient to reduce it to the factor space $\hat{M} = M/ R\times U(1)^2 $ by acting with the
Killing field $\eta$ associated with the azimuthal symmetry of the 2D sphere at infinity

\begin{eqnarray}
M_{ADM} =   {L\over 8} \int_{Arc(\infty)} \left[2i_\eta i_k \star d\xi - i_\eta i_\xi \star d k \right].
\end{eqnarray}
The integration is now performed over the semicircle representing the boundary of the two-dimensional factor space at infinity.
Using the Stokes' theorem the integral can be further expanded into a bulk term over $\hat{M}$ and an integral over the rest of the boundary
of the factor space, which is represented by the interval structure $I_n$

\begin{eqnarray}
M_{ADM} = {L\over 8} \int_{\hat{M}} \left[2d i_\eta i_k \star d\xi
- d i_\eta i_\xi \star d k \right] + {L\over 8}\sum_n \int_{I_n}
\left[2i_\eta i_k \star d\xi - i_\eta i_\xi \star d k \right] .
\end{eqnarray}

If we take into account the definitions of the intrinsic masses of the black holes $M_{H_i}$ and the bubbles $M_{B_j}$, and the fact that the
integral vanishes on the axis of rotation, we obtain

\begin{eqnarray}\label{M}
M_{ADM} =  \sum_{i} M_{H_i} + \sum_{j} M_{B_j} + {L\over 8} \int_{\hat{M}} \left[2d i_\eta i_k \star d\xi
- d i_\eta i_\xi \star d k \right],
\end{eqnarray}
where the index $i$ enumerates the horizons, and the index $j$ runs over the Kaluza-Klein bubbles. Let us consider the bulk integral and
use the Ricci-identity $d\star d \zeta=2\star R(\zeta)$, which applies for any Killing field $\zeta$

\begin{eqnarray}
 {L\over 8} \int_{\hat{M}} \left[ 2d i_\eta i_k \star d\xi - d i_\eta i_\xi \star d k \right] &=&
 {L\over 4} \int_{\hat{M}} \left[2i_\eta i_k \star R(\xi) -
i_\eta i_\xi \star R(k) \right].
\end{eqnarray}
We further apply the relation
\begin{eqnarray}
\star R(\zeta) = - 2e^{-2a\varphi} \left( -{2\over 3}i_{\zeta}F\wedge \star F + {1\over 3} F\wedge i_{\zeta}\star F \right),
\end{eqnarray}
which is also valid for any Killing field $\zeta$,  for the Killing fields $\xi$ and $k$. Considering the explicit form of the electromagnetic field
we obtain

\begin{eqnarray}
2i_\eta i_k \star R(\xi) - i_\eta i_\xi \star R(k) &=& 2\left( i_\xi F
\wedge i_\eta i_k e^{-2\alpha\varphi}\star F + i_\eta F \wedge i_\xi
i_k e^{-2\alpha\varphi}\star F\right) \\[2mm] \nonumber
&=& 2\left(dA_t \wedge d{\cal B}_\eta + dA_\phi \wedge d{\cal B}_\xi\right).
\end{eqnarray}

The bulk term can be simplified using the Stokes' theorem and taking into account that the integrals at infinity vanish, as well as the integrals
over the bubble intervals and over the axis of the Killing field $\eta$. Thus, we obtain

\begin{eqnarray}
 {L\over 2} \int_{\hat{M}}\left[ dA_t \wedge d{\cal B}_\eta + dA_\phi \wedge d{\cal B}_\xi\right]
 = - {L\over 2}\sum_i \int_{I_{H_i}}\left[ A_t d{\cal B}_\eta + A_\phi d{\cal B}_\xi\right].
\end{eqnarray}

We substitute this expression in equation $(\ref{M})$ and apply the identity for the horizon Komar mass $(\ref{M_Komar})$, which leads to

\begin{eqnarray}
M_{ADM} &=&   \sum_{i} M_{H_i} + \sum_{j} M_{B_j} - {L\over 2}\sum_i \int_{I_{H_i}}\left[ A_t d{\cal B}_\eta + A_\phi d{\cal B}_\xi\right] \\[2mm] \nonumber
&=&\sum_{i} M^{(0)}_{H_i} + \sum_{j} M^{(0)}_{B_j} - {L\over 2}\sum_i \int_{I_{H_i}}\left[ A_t d{\cal B}_\eta + \Omega_i A_\phi d{\cal B}_\eta\right] \\[2mm] \nonumber
&=&\sum_{i} M^{(0)}_{H_i} + \sum_{j} M^{(0)}_{B_j} + \sum_i\Sigma_{H_i}{\cal Q}_{H_i} \\[2mm] \nonumber
&=& \sum_{i}(1+\frac{1}{2}\sinh^2\gamma) M^{(0)}_{H_i} + \sum_{j} M^{(0)}_{B_j}.
\end{eqnarray}

We can also consider the relation of the total angular momentum to some local quantities by expanding the corresponding Komar integral over the factor space
$\hat{M}$. Thus, we obtain

\begin{eqnarray}
{\cal J} &=&   -{L\over 8} \int_{Arc(\infty)} i_\eta i_k \star d\eta = -{L\over 8} \sum_i\int_{I_{H_i}} i_\eta i_k \star d\eta - {L\over 4} \int_{\hat{M}} i_\eta i_k \star R(\eta)  \\[2mm] \nonumber
&=& \sum_i J_{H_i} - {L\over 2} \int_{\hat{M}} d A_\phi\wedge d{\cal B}_\eta  = \sum_i\left[J_{H_i} + {L\over 2} \int_{I_{H_i}}  A_\phi d{\cal B}_\eta\right] \\[2mm] \nonumber
&=& \cosh\gamma \sum_i J^{(0)}_{H_i},
\end{eqnarray}
where we have applied the identity ($\ref{J_Komar}$) for the horizon angular momenta, and the index $i$ again runs over the horizons.
In the case when a single horizon is present we can express the Smarr relation for the mass in the following form

\begin{eqnarray}
M_{ADM}  = \frac{1}{4\pi}\kappa_{H}A_{H} + 2\Omega {\cal J} + \sum_{j} M_{B_j} + \Sigma_{H}{\cal Q}_{H},
\end{eqnarray}
using this relation.

In a similar way we can derive a Smarr relation for the tension. Considering the corresponding Komar integral
\begin{eqnarray}
{\cal T} L &=&  {L\over 8} \int_{Arc(\infty)} \left[i_\eta i_k \star d\xi - 2i_\eta i_\xi \star d k \right] \\[2mm] \nonumber
&=&  \frac{1}{2}\sum_{i} M_{H_i} + 2\sum_{j} M_{B_j} + {L\over 4} \int_{\hat{M}} \left[i_\eta i_k \star R(\xi) -
2i_\eta i_\xi \star R(k) \right],
\end{eqnarray}
and using the identity
\begin{eqnarray}
i_\eta i_k \star R(\xi) - 2i_\eta i_\xi \star R(k) &=& 2 i_\eta F \wedge i_\xi i_k e^{-2\alpha\varphi}\star F
= 2 dA_\phi \wedge d{\cal B}_\xi,
\end{eqnarray}
as well as the expression for the Komar mass of each horizon, we obtain

\begin{eqnarray}
{\cal T} L &=&  \frac{1}{2}\sum_{i} M_{H_i} + 2\sum_{j} M^{(0)}_{B_j} - {L\over 2} \sum_i \int_{I_{H_i}} A_\phi d{\cal B}_\xi \\[2mm] \nonumber
 &=& \frac{1}{2}\sum_{i} M^{(0)}_{H_i} + 2\sum_{j} M^{(0)}_{B_j}
\end{eqnarray}
\noindent
Consequently, the Smarr relation for the tension coincides with that for the seed solution.

\subsection{Gyromagnetic ratio}

We can investigate the gyromagnetic ratio of the constructed solutions. It is defined as the constant of proportionality $g$ in the relation

\begin{eqnarray}
{\cal M} = g\frac{{\cal Q}{\cal J}}{2M_{ADM}},
\end{eqnarray}
between the magnetic moment, the angular momentum, the ADM mass and the electric charge of the solution. Since the magnetic moment and the angular momentum are not
independent, but related as ${\cal M} = \tanh\gamma {\cal J}$, the gyromagnetic ratio does not depend on the value of the angular momentum, but only on the
ratio between the ADM mass and the electric charge. We use the Smarr-like relation for the mass

\begin{eqnarray}
M_{ADM} = \sum_{i}(1+\frac{1}{2}\sinh^2\gamma) M^{(0)}_{H_i} + \sum_{j} M^{(0)}_{B_j},
\end{eqnarray}
and the expression for the local electric charge on each horizon

\begin{eqnarray}
Q_{H_i} = \sinh\gamma\cosh\gamma M^{(0)}_{H_i},
\end{eqnarray}
and also take into account that the total electric charge of the solution is a sum of the local charges on all of the horizons. Then, we can express the
gyromagnetic ratio in the form

\begin{eqnarray}
g = 2\cosh^{-2}\gamma\left( 1 + \frac{1}{2}\sinh^2\gamma + \frac{\sum_{j} M^{(0)}_{B_j}}{\sum_{i} M^{(0)}_{H_i}}\right),
\end{eqnarray}
where $M^{(0)}_{H_i}$ and $M^{(0)}_{B_j}$ are the Komar masses of the black holes and the bubbles in the seed solution. For large electric charges
$\gamma \rightarrow \infty$, the gyromagnetic ratio approaches the value $g=1$.

\section{Particular solutions}

\subsection{Charged rotating dilaton black string}

The most simple solution which belongs to the class we consider is the black string. It contains a single horizon and can be described by the interval
structure represented in fig. \ref{rodstr_BS}.  The rotating black string in vacuum is constructed by trivially embedding the Kerr black hole in 5D spacetime by adding
a compact extra dimension. By performing the Kaluza-Klein transformation on it we obtain its charged dilaton generalization. The solution is represented in
the most simple form in the Boyer-Lindquist coordinates, in which the metric is given by

\begin{eqnarray}
ds^2&=&-\Lambda^{-\frac{2}{3}}\,\frac{\Delta - a^2\sin^2\theta}{\Sigma}\left( dt + \omega d\phi \right)^2+ \Lambda^{\frac{1}{3}}\,
\left[ \frac{\Sigma\Delta\sin^2\theta}{\Delta - a^2\sin^2\theta} d\phi^2 + \frac{\Sigma}{\Delta}dr^2 + \Sigma d\theta^2 + d\psi^2\right],  \nonumber \\[2mm]
\omega&=& \frac{2mra\sin^2\theta}{\Delta-a^2\sin^2\theta}\cosh\gamma, \\[2mm] \nonumber
\Lambda &=& 1 + \frac{2mr}{\Sigma}\sinh^2\gamma, \\[2mm] \nonumber
\Delta &=& r^2-2mr + a^2, \quad~~~ \Sigma = r^2 + a^2\cos^2\theta, \nonumber
\end{eqnarray}
and the electromagnetic potentials are

\begin{eqnarray}
A_t &=& \Lambda^{-1}\sinh\gamma\cosh\gamma \frac{mr}{\Sigma}, \\ \nonumber
A_\phi &=& -\Lambda^{-1}\sinh\gamma\frac{mr}{\Sigma}a\sin^2\theta.
\end{eqnarray}

It is characterized by the real parameters $a \in (0,1)$, $m>0$, which correspond to the spin parameter of the vacuum seed solution $a= J^{(0)}/M^{(0)}$,
and its mass normalized to the length of the compact dimension $m=M^{(0)}/L$ . The length of the horizon interval in the interval structure $\Delta l = 2\sigma$
is connected to them by the relation $\sigma^2 = m^2-a^2$. The horizon is located at $r=r_H= m + \sqrt{m^2-a^2}$, and rotates with angular velocity

\begin{eqnarray}
\Omega = \frac{a\cosh^{-1}\gamma}{2m(\sigma + m)}.
\end{eqnarray}

The $ADM$ mass, angular momentum and electric charge of the solution are given by
\begin{eqnarray}
M_{ADM} &=& L\,m(1 + \frac{1}{2}\sinh^2\gamma),  \\ \nonumber
{\cal J} &=& L\cosh\gamma\, am, \\ \nonumber
{\cal Q} &=& L\sinh\gamma\cosh\gamma\, m,
\end{eqnarray}
and the gyromagnetic ratio is equal to $g= 1+\cosh^{-2}\gamma$.

We can obtain the horizon Komar mass and angular momentum by calculating the integral

\begin{eqnarray}
I= L\int_{I_H}A_\phi d {\cal B}_\xi =  -\frac{L}{2}\sinh\gamma \int_{I_H}A_\phi d\chi^{(0)}.
\end{eqnarray}

The twist 1-form for the vacuum seed solution on the horizon takes the form

\begin{equation}
d\chi^{(0)} = -\frac{\alpha(1+\alpha^2)(1-\alpha^2\cos^2\theta)\sin\theta}{(1+\alpha^2\cos^2\theta)^2}d\theta,
\end{equation}
where $\alpha = a/r_H$, and the restriction of the electromagnetic potential $A_\phi$ on the horizon is given by

\begin{equation}
A_\phi =  -\sinh\gamma\frac{m\alpha\sin^2\theta}{\beta^2 +\alpha^2\cos^2\theta},
\end{equation}
where we define the parameter $\beta^2 = 1 + \sinh^2\gamma (1+\alpha^2)$. Consequently, the integral is equal to\footnote{Instead of the mass and spin parameters $m$ and $a$, we can parameterize
the black string solution with the parameters $\sigma$ and $\alpha$. They are related as $m = \frac{\sigma(1+\alpha^2)}{1-\alpha^2}$ and
$a = \frac{2\sigma\alpha}{1-\alpha^2}$.}

\begin{eqnarray}
I= -L\sigma\frac{(1+\alpha^2)}{1-\alpha^2} + L\sigma\cosh^2\gamma \frac{(1+\alpha^2)^2}{\alpha\beta(1-\alpha^2)(\beta^2-1)}\left(2\beta\arctan\alpha
 - (1+\beta^2)\arctan\frac{\alpha}{\beta}\right). \nonumber
\end{eqnarray}
Taking into account the expression for the horizon Komar mass of the seed solution

\begin{equation}
M^{(0)}_H = Lm = L\sigma\frac{1+\alpha^2}{1-\alpha^2},
\end{equation}
we obtain for the horizon Komar mass

\begin{eqnarray}
M_H =L\sigma\cosh^2\gamma \frac{(1+\alpha^2)^2}{\alpha\beta(1-\alpha^2)(\beta^2-1)}\left(2\beta\arctan\alpha -
(1+\beta^2)\arctan\frac{\alpha}{\beta}\right).
\end{eqnarray}
The horizon angular momentum is calculated from $(\ref{J_Komar})$ using this result and the fact that on the horizon it is satisfied that
$d{\cal B}_\eta = -\Omega^{-1}_H d{\cal B}_\xi$. We obtain the following expression

\begin{eqnarray}
J_H &=&-\frac{L\sigma^2 (1+\alpha^2)}{\alpha(1-\alpha^2)}\cosh\gamma\bigg[1 -  \cosh^2\gamma\frac{(1+\alpha^2)^2}{\alpha\beta(1-\alpha^2)(\beta^2-1)}
\bigg(2\beta\arctan\alpha   \nonumber \\[2mm]
&-& (1+\beta^2)\arctan\frac{\alpha}{\beta}\bigg)\bigg].\nonumber
\end{eqnarray}

\subsection{Charged rotating black ring on Kaluza-Klein bubbles}

\begin{figure}[h]
\centering
\begin{minipage}[b]{0.45\linewidth}
\setlength{\tabcolsep}{ 0 pt }{\scriptsize\tt
\begin{tabular}{ cc }
\hspace{1.5cm}	 \includegraphics[width=4.2cm]{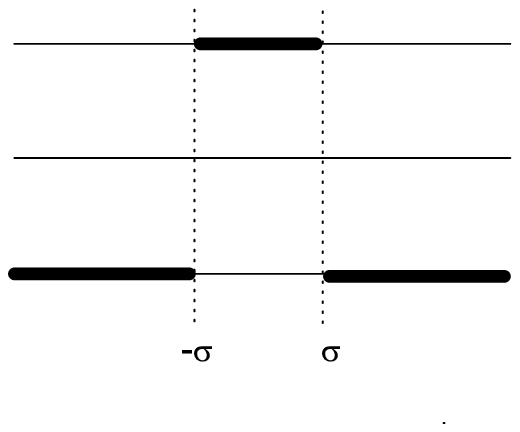}
                  \end{tabular}}
           \begin{picture}(0,0)(0,0)
\put(-112, -5){{\tiny(0,1,0)}}
\put(-76,48){{\tiny(1,$\Omega$,0)}}
\put(-36,-6){{\tiny(0,1,0)}}
\end{picture}
\end{minipage}
\quad
\begin{minipage}[b]{0.5\linewidth}
\setlength{\tabcolsep}{ 0 pt }{\scriptsize\tt
\begin{tabular}{ cc }
	 \includegraphics[width=6.2cm]{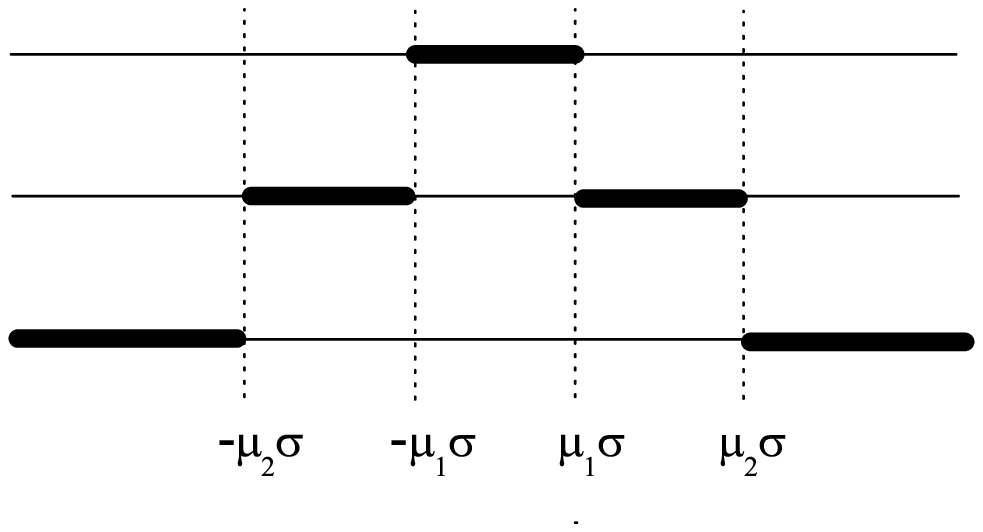}
                  \end{tabular}}
           \begin{picture}(0,0)(0,0)
\put(-170, -5){{\tiny(0,1,0)}}
\put(-132,20){{\tiny(0,0,1)}}
\put(-104, 46){{\tiny(1,$\Omega$,0)}}
\put(-74,20){{\tiny(0,0,1)}}
\put(-38,-6){{\tiny(0,1,0)}}
\end{picture}
\end{minipage}
\caption{\footnotesize{Interval structure of the rotating black string (left), and the rotating black ring on Kaluza-Klein bubbles (right).}}
\label{rodstr_BS}
\end{figure}

The charged rotating black ring on Kaluza-Klein bubbles is an example of a solution which contains a single horizon, but in addition also possesses
fixed point sets of the Killing field associated with the compact dimension.  The rotating black ring on Kaluza-Klein bubbles in vacuum was constructed
recently by means of solitonic techniques \cite{Nedkova:2010}. It is characterized by the interval structure represented in fig. \ref{rodstr_BS}.
If the solution is
balanced, it should contain bubbles with equal size. Then, it is described by 3 real parameters $\sigma >0$, $\mu_1 \in (0, 1)$ and $\mu_2 > 1$,
which are associated with the lengths of  the horizon and bubble intervals, and the angular momentum. Due to the balance conditions
the length of the compact dimension at infinity is not an  independent parameter, but is defined by the relation

\begin{eqnarray}
L = 8\pi\sigma\frac{\mu_1\mu_2(\mu_2 - \mu_1)}{(\mu_1\mu_2 - 1)(\mu_2 + \mu_1)}\sqrt{\mu_2^2 - 1}.
\end{eqnarray}

Applying the Kaluza-Klein transformation we obtain the metric of the charged dilaton generalization in the form
\begin{eqnarray}
ds^2 &=& -\Lambda^{-\frac{2}{3}}\frac{W_1}{W_2}\left(dt + \cosh\gamma\omega d\phi \right)^2 + \Lambda^{\frac{1}{3}}\bigg[\frac{W_2}{W_1}\rho^2
\frac{e^{2\widetilde{U}_{-\mu_2\sigma}}e^{2\widetilde{U}_{\mu_1\sigma}}}{e^{2\widetilde{U}_{-\mu_1\sigma}}
e^{2\widetilde{U}_{\mu_2\sigma}}}d\phi^2 + \frac{e^{2\widetilde{U}_{-\mu_1\sigma}}e^{2\widetilde{U}_{\mu_2\sigma}}}{e^{2\widetilde{U}_{-\mu_2\sigma}}
e^{2\widetilde{U}_{\mu_1\sigma}}}d\psi^2   \nonumber  \\[2mm] \nonumber
&+& Y\left(d\rho^2 + dz^2\right)\bigg], \\[2mm] \nonumber
\Lambda &=& 1 + \sinh^2\gamma \frac{W_2-W_1}{W_2}, \\[2mm] \nonumber
\widetilde{U}_{c}&=&\frac{1}{2}\ln\left[\sqrt{\rho^2 + (z-c)^2}+(z-c)\right],
\end{eqnarray}
where $\rho$ and $z$ are Weyl coordinates, and the metric functions $W_1$, $W_2$, $\omega$ and $Y$ depend only on them. Due to their complexity, we provide
their explicit form in the Appendix.

The electromagnetic potentials are

\begin{eqnarray}
A_t &=& \frac{1}{2}\Lambda^{-1}\sinh\gamma\cosh\gamma \frac{W_2-W_1}{W_2}, \\ \nonumber
A_\phi &=& -\frac{1}{2}\Lambda^{-1}\sinh\gamma \frac{W_1}{W_2}\omega.
\end{eqnarray}

We obtain for the $ADM$ mass, the angular momentum and the electric charge of the solution the expressions
\begin{eqnarray}
M_{ADM} &=&  \frac{L}{2}\sigma(\mu_2-\mu_1)\frac{(\mu_1\mu_2 + 1)}{(\mu_1\mu_2 - 1)} + \frac{L}{2}\sinh^2\gamma\,\sigma\frac{(\mu_2-\mu_1)}{(\mu_1\mu_2 - 1)},  \\ \nonumber
{\cal J} &=& L \cosh\gamma\, \sigma^2~\frac{\mu_1\mu_2(\mu_2 - \mu_1)}{(\mu_1\mu_2 - 1)^2}\sqrt{(1-\mu^2_1)(\mu^2_2 - 1)}, \\ \nonumber
{\cal Q} &=& L\sinh\gamma\cosh\gamma\, \sigma\frac{(\mu_2-\mu_1)}{(\mu_1\mu_2 - 1)}.
\end{eqnarray}

The intrinsic masses of the black hole and the bubbles for the vacuum seed solution are given by

\begin{eqnarray}
M^{(0)}_{H}&=& L\sigma\frac{(\mu_2-\mu_1)}{(\mu_1\mu_2 - 1)}, \\ \nonumber
M^{(0)}_{B_j}&=& \frac{L}{4}\sigma(\mu_2-\mu_1),  \quad~~~ j=1,2.
\end{eqnarray}

Using these expressions we can calculate the gyromagnetic ratio of the charged solution

\begin{eqnarray}
g = 1 + \cosh^{-2}\gamma\,\mu_1\mu_2.
\end{eqnarray}

The horizon angular velocity and area are

\begin{eqnarray}
\Omega &=& \frac{1}{2\sigma\cosh\gamma}\frac{\mu_1\mu_2 - 1}{\mu_1(\mu_2 - \mu_1)}\sqrt{\frac{(1-\mu^2_1)}{(\mu^2_2 - 1)}}, \nonumber \\[2mm]
A_{H} &=& 32\pi L \cosh\gamma \sigma^2 \frac{\mu_1^3\mu_2(\mu_2-\mu_1)(\mu_2^2-1)}{(\mu_1\mu_2- 1)^2(\mu_1+\mu_2)^2}.
\end{eqnarray}

The horizon Komar mass and angular momentum are obtained by calculating the integral

\begin{eqnarray}\label{Int_Komar2}
I= L\int_{I_H}A_\phi d {\cal B}_\xi =  -\frac{L}{2}\sinh\gamma \int_{I_H}A_\phi d\chi^{(0)},
\end{eqnarray}
and using the relation $(\ref{M_Komar})$. The twist potential of the seed solution possesses the following form on the horizon

\begin{eqnarray}
\chi^{(0)} = \beta \frac{y(1+\mu_1)(1+b^2)}{(1+\mu_2)(1+b^2y^2)},
\end{eqnarray}
where $y = z/\sigma$, and
\begin{eqnarray}
b^2=\frac{(1-\mu_1^2)(\mu_2^2-y^2)}{(\mu_2^2-1)(\mu_1^2-y^2)}, \quad~~~
\beta =  -\sqrt{ \frac{(1-\mu_1)(1+\mu_2)}{(\mu_2-1)(1+\mu_1)}}. \\ \nonumber
\end{eqnarray}

\noindent
The restriction of the electromagnetic potential $A_\phi$ on the horizon is given by

\begin{eqnarray}
A_\phi = -\frac{1}{2}\frac{\sinh\gamma}{\Omega^{(0)}}\frac{(1-y^2)b^2}{1+b^2y^2 + \sinh^2\gamma(1+b^2)}. \\ \nonumber
\end{eqnarray}

\noindent
Consequently, the integral $(\ref{Int_Komar2})$ is equal to

\begin{eqnarray}
I &=& -L\sigma \frac{(\mu_2-\mu_1)}{\mu_1\mu_2 - 1}\left[ 1 -  \frac{\mu_1^2}{\tanh^2\gamma} \left( \frac{2}{Z}\arctan{Z}
- \frac{F^2 + Z^2}{FZ^2}\arctan{F}\right)\right] \nonumber \\[4mm]
&=& -M^{(0)}_H + M_H, \nonumber \\[4mm]
F^2 &=& \frac{\mu_1^2(1-\mu_1^2)}{\mu_1^2(\mu_2^2-1) + \sinh^2\gamma(\mu_2^2-\mu_1^2)}, \quad~~~ Z^2 = \frac{1-\mu_1^2}{\mu_2^2-1},  \\ \nonumber
\end{eqnarray}

\noindent
and we obtain for the horizon Komar mass

\begin{eqnarray}
M_H &=& L\sigma \frac{\mu_1^2(\mu_2-\mu_1)}{\tanh^2\gamma(\mu_1\mu_2 - 1)} \left( \frac{2}{Z}\arctan{Z}-\frac{F^2 + Z^2}{FZ^2}\arctan{F} \right).
 \\ \nonumber
\end{eqnarray}

The horizon angular momentum is calculated in a similar way using the relation $(\ref{J_Komar})$, and the evaluated integral $(\ref{Int_Komar2})$

\begin{eqnarray}
J_H &=& -L\cosh\gamma\mu_1^2\sigma^2\frac{(\mu_2-\mu_1)}{\mu_1\mu_2 - 1}\sqrt{\frac{\mu_2^2-1}{1-\mu_1^2}}\bigg[1+\frac{\mu_1(\mu_2-\mu_1)}{\tanh^2\gamma(\mu_1\mu_2
- 1)} \bigg( \frac{F^2 + Z^2}{FZ^2}\arctan{F}  \nonumber \\[3mm]
&-& \frac{2}{Z}\arctan{Z}\bigg)\bigg].\nonumber
\end{eqnarray}

\subsection{Charged rotating black holes on a Kaluza-Klein bubble}

\begin{figure}[h]
\centering
\setlength{\tabcolsep}{ 0 pt }{\scriptsize\tt
\begin{tabular}{ cc }
	 \includegraphics[width=6.2cm]{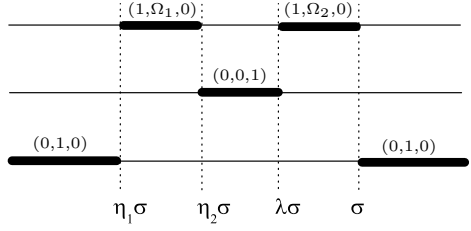}
                  \end{tabular}}
           \begin{picture}(0,0)(0,0)
\put(-170, -5){{\tiny(0,1,0)}}
\put(-134,46){{\tiny(1,$\Omega_1$,0)}}
\put(-102,21){{\tiny(0,0,1)}}
\put(-75,46){{\tiny(1,$\Omega_2$,0)}}
\put(-38,-6){{\tiny(0,1,0)}}
\end{picture}
\caption{\footnotesize{Interval structure of the rotating black holes on a Kaluza-Klein bubble.}}
\label{rodstr_BHB}
\end{figure}

As a next example, we consider a multi-horizon solution, which describes a superposition of two rotating black holes and a Kaluza-Klein bubble.
The vacuum solution was constructed in \cite{Tomizawa:2007} using the inverse scattering method, and it is described by the interval structure presented
in fig. \ref{rodstr_BHB}.
The solution is characterized by 4 parameters $\eta_1 < \eta_2 <-1 <\lambda <1$, and $\sigma >0$, which are related to the lengths of the horizon and bubble
intervals in the interval structure, and the angular momentum of the solution. The solution is free of conical singularities, if the length of the
Kaluza-Klein circle at infinity is related to the solution parameters as

\begin{eqnarray}
L = 4\pi\sigma\frac{(\eta_1+1)((\eta_1-1)^2+(\lambda+1)(\eta_2-1))}
                           {(\eta_1-1)((\eta_1+1)^2+(\lambda+1)(\eta_2+1))}
             \sqrt{\frac{(\lambda-\eta_1)(\lambda-\eta_2)(\eta_2+1)}{\eta_2-1}}. \\ \nonumber
\end{eqnarray}

We apply the Kaluza-Klein transformation on this solution and obtain a charged dilaton solution with metric
\begin{eqnarray}
ds^2 &=& -\Lambda^{-\frac{2}{3}}\frac{W_1}{W_2}\frac{e^{2\widetilde{U}_{\eta_2\sigma}}}{e^{2\widetilde{U}_{\eta_1\sigma}}}\left(dt + \cosh\gamma\omega d\phi \right)^2
+ \Lambda^{\frac{1}{3}}\bigg[\frac{W_2}{W_1}\rho^2 \frac{e^{2\widetilde{U}_{\eta_1\sigma}}}{e^{2\widetilde{U}_{\lambda\sigma}}}d\phi^2 +
\frac{e^{2\widetilde{U}_{\lambda\sigma}}}{e^{2\widetilde{U}_{\eta_2\sigma}}}d\psi^2   \nonumber  \\[2mm] \nonumber
&+& Y\left(d\rho^2 + dz^2\right)\bigg], \\ \nonumber
\Lambda &=& 1 + \sinh^2\gamma \frac{W_2-W_1}{W_2}, \\[2mm] \nonumber
\widetilde{U}_{c}&=&\frac{1}{2}\ln\left[\sqrt{\rho^2 + (z-c)^2}+(z-c)\right],
\end{eqnarray}
where $\rho$ and $z$ are Weyl coordinates, and the metric functions $W_1$, $W_2$, $\omega$ and $Y$ depend only on them. Similar to the previous solution,
the metric functions are given by complicated expressions. Therefore, we provide their explicit form in the Appendix.

The electromagnetic potentials are

\begin{eqnarray}
A_t &=& \frac{1}{2}\Lambda^{-1}\sinh\gamma\cosh\gamma \frac{W_2-W_1}{W_2}, \\ \nonumber
A_\phi &=& -\frac{1}{2}\Lambda^{-1}\sinh\gamma \frac{W_1}{W_2}\omega.
\end{eqnarray}

The black hole horizons rotate with angular velocities
\begin{eqnarray}
\Omega_1 &=& \frac{-\beta(1-\beta^2)\cosh^{-1}\gamma}
             {[1-\eta_1+\beta^2(1+\eta_1)]^2\sigma}, \quad~~~ \beta^2 = -\frac{(1+\lambda)(1 +\eta_2)}{(1 + \eta_1)^2}, \nonumber \\[2mm]
\Omega_2 &=& \frac{-\beta(1-\beta^2)((1-\eta_2)(1-\lambda)+(1-\eta_1)^2)\cosh^{-1}\gamma}
                 {4[(1-\eta_1)^2+\beta^2(1-\eta_2)(1-\lambda)]\sigma}.
\end{eqnarray}

The $ADM$ mass, the angular momentum and the electric charge of the solution possess the following form

\begin{eqnarray}
M_{ADM}&=&\frac{L}{4}\frac{(4-2\eta_1+\eta_2+\lambda+\beta^2(4+2\eta_1-\eta_2-\lambda))\sigma}{1-\beta^2} \nonumber \\[2mm]
&+& \frac{L}{4}\sinh^2\gamma\frac{(\eta_2 - \eta_1 +2 - \beta^2 (\eta_2 -\eta_1 -2))\sigma}{1-\beta^2} ,  \nonumber \\[2mm]
{\cal J} &=& -\cosh\gamma\frac{2L\beta\sigma^2(2-2\eta_1+\eta_2+\lambda+\beta^2(2+2\eta_1-\eta_2-\lambda))}{(1-\beta^2)^2}, \nonumber \\[2mm]
{\cal Q} &=& \frac{L}{2}\sinh\gamma\cosh\gamma\frac{(\eta_2 - \eta_1 + 2 - \beta^2 (\eta_2 -\eta_1 - 2))\sigma}{1-\beta^2}.
\end{eqnarray}

We calculate the horizon and bubble Komar masses for the vacuum seed solution, and obtain
\begin{eqnarray}
M^{(0)}_{H_1} &=& \frac{L}{2}\frac{\sigma(\eta_2-\eta_1)}{(1-\beta^2)}\frac{\left[ 1 +\beta^2\frac{(1+\eta_1)}{(1-\eta_1)}\right]
\left[ 1 -\beta^2\frac{(1-\eta_2)(1+\eta_1)}{(1-\eta_1)(1+\eta_2)}\right]}
{\left[ 1 -\frac{(1-\eta_2)(1+\lambda)}{(1-\eta_1)^2}\right]}, \nonumber \\[3mm]
M^{(0)}_{H_2} &=& \frac{L\sigma}{(1-\beta^2)}\frac{\left[ 1 +\beta^2\frac{(1-\eta_2)(1-\lambda)}{(1-\eta_1)^2}\right]}
{\left[ 1 -\frac{(1-\eta_2)(1+\lambda)}{(1-\eta_1)^2}\right]}, \nonumber \\[3mm]
M^{(0)}_{B} &=& \frac{L}{4}(\lambda-\eta_2)\sigma,
\end{eqnarray}
where we denote by $M^{(0)}_{H_1}$ the Komar mass of the horizon located at $\eta_1\sigma < z < \eta_2\sigma$, and by $M^{(0)}_{H_2}$ the Komar mass of
the horizon located at $\lambda\sigma < z < \sigma$. Using these expressions and the relation $(\ref{Q})$, we can calculate the local charges on the
horizons

\begin{eqnarray}
{\cal Q}_{H_1} &=& \frac{L}{2}\sinh\gamma\cosh\gamma\frac{\sigma(\eta_2-\eta_1)}{(1-\beta^2)}\frac{\left[ 1 +\beta^2\frac{(1+\eta_1)}{(1-\eta_1)}\right]
\left[ 1 -\beta^2\frac{(1-\eta_2)(1+\eta_1)}{(1-\eta_1)(1+\eta_2)}\right]}
{\left[ 1 -\frac{(1-\eta_2)(1+\lambda)}{(1-\eta_1)^2}\right]}, \nonumber \\[3mm]
{\cal Q}_{H_2} &=& \sinh\gamma\cosh\gamma\frac{L\sigma}{(1-\beta^2)}\frac{\left[ 1 +\beta^2\frac{(1-\eta_2)(1-\lambda)}{(1-\eta_1)^2}\right]}
{\left[ 1 -\frac{(1-\eta_2)(1+\lambda)}{(1-\eta_1)^2}\right]}.
\end{eqnarray}

\section{Conclusion}

We construct a class of stationary and axisymmetric solutions to the 5D Einstein-Maxwell-dilaton equations, which describe sequences of charged rotating black
objects and Kaluza-Klein bubbles. The solutions are obtained for the special value of the dilaton coupling constant $\alpha=\sqrt{8/3}$ by uplifting a 5D vacuum solution with the same interval structure to six dimensions, performing a boost and a subsequent circle reduction. We investigate the general relation of their physical properties to the properties of the corresponding vacuum solutions. Certain relations are obtained between the twist potentials and the electromagnetic potentials for the charged and the vacuum cases, which lead to relations between some local properties of the solutions. The horizon Komar masses and angular momenta are modified with respect to the vacuum ones by a term reflecting the interaction with the electromagnetic field. On the other hand, the electric charge associated with each horizon is proportional to the corresponding horizon Komar mass in the vacuum solution. Although our class of solutions possess Kaluza-Klein asymptotics, these relations are also valid for asymptotically flat solutions obtained by the same procedure,
up to certain normalization factors. We further derive the Smarr-like relations and obtain an expression for the gyromagnetic ratio.

We investigate three exact solutions as particular cases. The first one represents the charged rotating dilaton black string. It is the most simple solution
belonging to the class we consider, since it contains only a single horizon and no fixed point sets of the Killing vector associated with the compact
dimension. As more complicated examples we investigate the charged rotating black ring sitting on Kaluza-Klein bubbles, and a multi-horizon solution describing a superposition of two rotating charged black holes and a Kaluza-Klein bubble.

\section{Appendix}

\subsection{Rotating black ring on Kaluza-Klein bubbles}

The metric of the rotating black ring on Kaluza-Klein bubbles in vacuum is given by
\begin{eqnarray}
ds^2 &=& -\frac{W_1}{W_2}\left(dt + \omega d\phi \right)^2 + \frac{W_2}{W_1}\rho^2\frac{e^{2\widetilde{U}_{-\mu_2\sigma}}
e^{2\widetilde{U}_{\mu_1\sigma}}}{e^{2\widetilde{U}_{-\mu_1\sigma}}
e^{2\widetilde{U}_{\mu_2\sigma}}}d\phi^2 + \frac{e^{2\widetilde{U}_{-\mu_1\sigma}}e^{2\widetilde{U}_{\mu_2\sigma}}}{e^{2\widetilde{U}_{-\mu_2\sigma}}
e^{2\widetilde{U}_{\mu_1\sigma}}}d\psi^2   \nonumber  \\[2mm] \nonumber
&+& Y\left(d\rho^2 + dz^2\right). \\ \nonumber
\end{eqnarray}
\noindent
The metric functions $W_1$, $W_2$ and $\omega$ have the form
\begin{eqnarray}\label{metricfunc}
W_1&=&\left[(R_{-\sigma}+R_{\sigma})^2-4\sigma^2\right] (1+ a b)^2  + \left[(R_{-\sigma}-R_{\sigma})^2-4\sigma^2\right](a-b)^2  , \nonumber \\[2mm]
W_2&=&\left[(R_{-\sigma}+R_{\sigma}+2\sigma)+(R_{-\sigma}+R_{\sigma}-2\sigma)a b \right]^2 \nonumber \\[2mm]
 &+& \left[(R_{-\sigma}-R_{\sigma}-2\sigma)a - (R_{-\sigma}-R_{\sigma}+2\sigma)b \right]^2 , \\ \nonumber  \\
\omega&=& - \frac{e^{\widetilde{U}_{-\mu_2\sigma}}e^{\widetilde{U}_{\mu_1\sigma}}}{e^{\widetilde{U}_{-\mu_1\sigma}}
e^{\widetilde{U}_{\mu_2\sigma}}}\frac{\hat{\omega}}{W_1}-\frac{4\sigma\alpha}{1 - \alpha^2}, \nonumber \\ \nonumber \\
{\hat \omega}&=& [(R_{-\sigma} + R_{\sigma})^2-4\sigma^2](1+a b)\left[(R_{-\sigma}-R_{\sigma} + 2\sigma)b +
(R_{-\sigma}-R_{\sigma} - 2\sigma)a\right] \nonumber \\[2mm]
&-& [(R_{-\sigma} - R_{\sigma})^2-4\sigma^2](b-a)
\left[(R_{-\sigma} + R_{\sigma} + 2\sigma) - (R_{-\sigma} + R_{\sigma} - 2\sigma)ab\right],  \nonumber
\end{eqnarray}
where the following functions are used

\begin{eqnarray} \label{ab}
a &=& \alpha
{e^{2U_{\sigma}} + e^{2{\widetilde U}_{-\mu_1\sigma}} \over e^{{\widetilde U}_{-\mu_1\sigma}} }
{e^{{\widetilde U}_{-\mu_2\sigma}} \over e^{2U_{\sigma}} + e^{2{\widetilde U}_{-\mu_2\sigma}}}
{e^{2U_{\sigma}} + e^{2{\widetilde U}_{\mu_2\sigma}} \over e^{{\widetilde U}_{\mu_2\sigma}}}
{e^{{\widetilde U}_{\mu_1\sigma}} \over e^{2U_{\sigma}} + e^{2{\widetilde U}_{\mu_1\sigma}}} , \\ \nonumber \\ \nonumber \\ \nonumber
b &=& \beta
{e^{2U_{-\sigma}} + e^{2{\widetilde U}_{-\mu_2\sigma}} \over e^{{\widetilde U}_{-\mu_2\sigma}} }
{e^{{\widetilde U}_{-\mu_1\sigma}} \over e^{2U_{-\sigma}} + e^{2{\widetilde U}_{-\mu_1\sigma}}}
{e^{2U_{-\sigma}} + e^{2{\widetilde U}_{\mu_1\sigma}} \over e^{{\widetilde U}_{\mu_1\sigma}}}
{e^{{\widetilde U}_{\mu_2\sigma}} \over e^{2U_{-\sigma}} + e^{2{\widetilde U}_{\mu_2\sigma}}} , \\ \nonumber \\ \nonumber
R_{c}  &=& \sqrt{\rho^2 + (z-c)^2},  \\ \nonumber \\ \nonumber
\widetilde{U}_{c}&=&\frac{1}{2}\ln\left[R_{c}+(z-c)\right],  \quad  U_{c}=\frac{1}{2}\ln\left[R_{c}-(z-c)\right].
\end{eqnarray}

The constants $\alpha$ and $\beta$ are connected with the solution parameters as
\begin{eqnarray}\label{beta}
\alpha = -\beta = \sqrt{\frac{(1-\mu_1)(1 +\mu_2)}{(\mu_2-1)(1 + \mu_1)}}.
\end{eqnarray}

The remaining metric function $Y$ is given by
\begin{eqnarray}
Y &=& \frac{ W_2}{16(1 - \alpha^2)^2R_\sigma R_{-\sigma}}\sqrt{\frac{Y_{\sigma, -\mu_2\sigma}Y_{\sigma, \mu_1\sigma}Y_{-\sigma, -\mu_1\sigma}
Y_{-\sigma,  \mu_2\sigma}}{Y_{\sigma, -\mu_1\sigma}Y_{\sigma,  \mu_2\sigma}Y_{-\sigma, -\mu_2\sigma}Y_{-\sigma, \mu_1\sigma}}}\times \\ \nonumber \\ \nonumber
 &&\frac{Y_{-\mu_2\sigma, -\mu_1\sigma}Y_{-\mu_2\sigma, \mu_2\sigma}Y_{-\mu_1\sigma, \mu_1\sigma}Y_{\mu_1\sigma, \mu_2\sigma}}
 {R_{-\mu_2\sigma} R_{-\mu_1\sigma} R_{\mu_1\sigma} R_{\mu_2\sigma}Y_{-\mu_2\sigma, \mu_1\sigma}Y_{-\mu_1\sigma, \mu_2\sigma}}
\frac{e^{2\widetilde{U}_{-\mu_2\sigma}}e^{2\widetilde{U}_{\mu_1\sigma}}}{e^{2\widetilde{U}_{-\mu_1\sigma}}
e^{2\widetilde{U}_{\mu_2\sigma}}},
\end{eqnarray}
\noindent
where we define the notation
\begin{eqnarray}
Y_{cd}&=& R_cR_d+(z-c)(z-d)+\rho^2.
\end{eqnarray}

We consider only balanced solutions, i.e. solutions  which are free of conical singularities. If conical singularities are allowed, the metric possesses
more general form (see \cite{Nedkova:2010}).

\subsection{Rotating black holes on a Kaluza-Klein bubble}

The metric of the rotating black holes on Kaluza-Klein bubble in vacuum is given by
\begin{eqnarray}
ds^2 &=& -\frac{W_1}{W_2}\frac{e^{2\widetilde{U}_{\eta_2\sigma}}}{e^{2\widetilde{U}_{\eta_1\sigma}}}\left(dt + \omega d\phi \right)^2
+ \frac{W_2}{W_1}\rho^2 \frac{e^{2\widetilde{U}_{\eta_1\sigma}}}{e^{2\widetilde{U}_{\lambda\sigma}}}d\phi^2 +
\frac{e^{2\widetilde{U}_{\lambda\sigma}}}{e^{2\widetilde{U}_{\eta_2\sigma}}}d\psi^2   \nonumber  \\[2mm] \nonumber
&+& Y\left(d\rho^2 + dz^2\right). \\ \nonumber
\end{eqnarray}
\noindent
The metric functions $W_1$, $W_2$ and $\omega$ have the form
\begin{eqnarray}\label{metricfunc}
W_1&=&\left[(R_{-\sigma}+R_{\sigma})^2-4\sigma^2\right] (1+ a b)^2  + \left[(R_{-\sigma}-R_{\sigma})^2-4\sigma^2\right](a-b)^2  , \nonumber \\[2mm]
W_2&=&\left[(R_{-\sigma}+R_{\sigma}+2\sigma)+(R_{-\sigma}+R_{\sigma}-2\sigma)a b \right]^2 \nonumber \\[2mm]
 &+& \left[(R_{-\sigma}-R_{\sigma}-2\sigma)a - (R_{-\sigma}-R_{\sigma}+2\sigma)b \right]^2 , \\ \nonumber  \\
\omega&=&-\frac{e^{2\widetilde{U}_{\eta_1\sigma}}}
{e^{\widetilde{U}_{\lambda\sigma}}e^{\widetilde{U}_{\eta_2\sigma}}}\frac{\hat{\omega}}{W_1}-\frac{4\sigma\alpha}{1 - \alpha^2}, \nonumber \\ \nonumber \\
{\hat \omega}&=& [(R_{-\sigma} + R_{\sigma})^2-4\sigma^2](1+a b)\left[(R_{-\sigma}-R_{\sigma} + 2\sigma)b +
(R_{-\sigma}-R_{\sigma} - 2\sigma)a\right] \nonumber \\[2mm]
&-& [(R_{-\sigma} - R_{\sigma})^2-4\sigma^2](b-a)
\left[(R_{-\sigma} + R_{\sigma} + 2\sigma) - (R_{-\sigma} + R_{\sigma} - 2\sigma)ab\right],  \nonumber
\end{eqnarray}
where the following functions are used

\begin{eqnarray} \label{ab}
a &=& \alpha
{e^{2U_{\sigma}} + e^{2{\widetilde U}_{\lambda\sigma}} \over e^{{\widetilde U}_{\lambda\sigma}} }
{e^{2U_{\sigma}} + e^{2{\widetilde U}_{\eta_2\sigma}} \over e^{{\widetilde U}_{\eta_2\sigma}}}
\left({e^{{\widetilde U}_{\eta_1\sigma}} \over e^{2U_{\sigma}} + e^{2{\widetilde U}_{\eta_1\sigma}}}\right)^2 , \\ \nonumber \\ \nonumber \\ \nonumber
b &=& \beta
{e^{{\widetilde U}_{\lambda\sigma}} \over e^{2U_{-\sigma}} + e^{2{\widetilde U}_{\lambda\sigma}}}
{e^{{\widetilde U}_{\eta_2\sigma}} \over e^{2U_{-\sigma}} + e^{2{\widetilde U}_{\eta_2\sigma}}}
\left({e^{2U_{-\sigma}} + e^{2{\widetilde U}_{\eta_1\sigma}} \over e^{{\widetilde U}_{\eta_1\sigma}}}\right)^2 , \\ \nonumber \\ \nonumber
R_{c}  &=& \sqrt{\rho^2 + (z-c)^2},  \\ \nonumber \\ \nonumber
\widetilde{U}_{c}&=&\frac{1}{2}\ln\left[R_{c}+(z-c)\right],  \quad  U_{c}=\frac{1}{2}\ln\left[R_{c}-(z-c)\right].
\end{eqnarray}

The constants $\alpha$ and $\beta$ are connected with the solution parameters as
\begin{eqnarray}\label{beta}
\alpha = -\beta = \sqrt{-\frac{(1+\lambda)(1 +\eta_2)}{(1 + \eta_1)^2}}.
\end{eqnarray}

The remaining metric function $Y$ is given by

\begin{eqnarray}
Y &=& \frac{ W_2}{(1 - \alpha^2)^2R_\sigma R_{-\sigma}R_{\lambda\sigma}R_{\eta_1\sigma}R_{\eta_2\sigma}}\sqrt{\frac{Y_{-\sigma, \lambda\sigma}
Y_{-\sigma, \eta_2\sigma}Y_{\lambda\sigma,  \eta_1\sigma}Y_{\lambda\sigma,  \eta_2\sigma}Y_{\eta_1\sigma,  \eta_2\sigma}}
{Y_{\sigma, \lambda\sigma}Y_{\sigma, \eta_2\sigma}}}\times \\ \nonumber \\ \nonumber
 &&\frac{Y_{\sigma, \eta_1\sigma}}{Y_{-\sigma, \eta_1\sigma}}
\frac{e^{2\widetilde{U}_{\eta_1\sigma}}}{e^{2\widetilde{U}_{\lambda\sigma}}},
\end{eqnarray}
\noindent
where we define the notation
\begin{eqnarray}
Y_{cd}&=& R_cR_d+(z-c)(z-d)+\rho^2.
\end{eqnarray}

\section*{Acknowledgements}
We would like to thank Jutta Kunz for valuable discussions. The financial support by the DFG Research Training Group 1620 ``Models of Gravity'' is gratefully acknowledged. P.N. is partially supported by the Bulgarian NSF Grant $\textsl{DFNI T02/6}$.

\end{document}